# Resonant soft X-ray scattering reveals hierarchical structure in a multi-component vapor-deposited glass.

*Camille Bishop,[1] Thomas J. Ferron,[1] Marie E. Fiori,[2] Connor G. Bischak,[3] Eliot Gann,[1,4] Cherno Jaye,[4] Mark D. Ediger,[2] Dean M. DeLongchamp[1]*

1. National Institute of Standards and Technology, 100 Bureau Drive, Gaithersburg MD 20899, USA.

2. University of Wisconsin – Madison, Department of Chemistry, 1101 University Ave, Madison WI 53710, USA.

3. University of Utah, Department of Chemistry, 315 S 1400 E, Salt Lake City UT 84112, USA.

4. **Brookhaven National Laboratory, 98 Rochester St, Upton NY 11973, USA.**

**Abstract:**

Multi-phase vapor-deposited glasses are an important class of materials for organic electronics, particularly organic photovoltaics and thermoelectrics. These blends are frequently regarded as molecular alloys and there have been few studies of their structure at nanometer scales. Here we show that a co-deposited system of TPD (N,N'-Bis(3-methylphenyl)-N,N'-diphenylbenzidine) and DO37 (Disperse Orange 37), two small molecule glass-formers, separates into compositionally distinct phases with a domain size and spacing that depends on substrate temperature during deposition. Domains rich in one of the two components become larger and more pure at higher deposition temperatures. We use resonant soft X-ray scattering (RSoXS) complemented with Atomic Force Microscopy (AFM) and photo-induced force microscopy (PiFM) to measure the phase separation, topography, and purity of the deposited films. A forward-simulation approach to RSoXS analysis, the National Institute of Standards and Technology (NIST) RSoXS Simulation Suite (NRSS), is used with models developed from AFM images to evaluate the energy dependence of scattering across multiple length scales and interpret the RSoXS with respect to structure within the films. We find that the RSoXS is sensitive to a hidden length scale of phase separation that is not apparent from the AFM characterization alone. We demonstrate that vacuum scattering, which is often ignored in RSoXS analysis, contributes significantly to the features and energy dependence of the RSoXS pattern, and then illustrate how to properly account for vacuum scattering to analyze films with significant roughness. We then use this analysis framework to understand structure development mechanisms that occur during vapor deposition of a TPD-DO37 co-deposited glass with results that outline paths to tune morphology in multi-component materials.

*Introduction.*

Multi-component organic films, in both crystalline and glassy states, form critical functional layers in a wide variety of established and emerging organic electronics technologies. Glasses are ideal for displays due to their macroscopic homogeneity, compositional flexibility, and tunability. Organic light-emitting diodes (OLEDs) frequently have small concentrations (≈0.06 mass fraction) of dopants in a glassy host.[1,2] Although aggregation or compositional phase separation in such systems can be deleterious for emissive devices due to excited state quenching, it can be exploited in other applications. Well-controlled separation is desirable in certain applications such as thermoelectrics[3,4] and bulk heterojunction (BHJ) organic photovoltaics. In BHJ materials, a donor and acceptor are mixed, increasing the interfacial area of available heterojunction and decreasing necessary exciton transport distances relative to traditional planar geometries.[5,6] Because organic photovoltaics are a promising clean energy technology, extensive effort has been made to control morphology in both solution-processed and vapor-deposited BHJs.[7]

Physical vapor deposition (PVD) is one route to preparing out-of-equilibrium amorphous materials with enhanced stability and molecular packing that is inaccessible from bulk processing. With this technique, molecules are thermally evaporated in a high vacuum chamber where they deposit on an often temperature-controlled substrate. PVD produces continuous, conformal, macroscopically homogenous films with precisely controlled thickness without the need for solvent, allowing them to be manufactured in stacks without perturbing, or being perturbed by, the layers below.[8,9] PVD can make exceptional glasses due to a free interface-mediated equilibration mechanism; molecules at the free surface experience enhanced mobility, allowing them to partially equilibrate before becoming a bulk solid. This enhanced equilibration can lead to increased density,[10,11] thermal and kinetic stability,[12–14] and molecular orientation[2,15,16] and packing structures[17–19] that are not accessible through cooling a bulk liquid alone. Finally, as the molecular packing can be finely controlled through both substrate temperature and deposition rate,[20] a wide range of materials can be made by choice of easily controlled processing parameters. All of these qualities make vapor-deposited materials ideal for employment in high-performance, stable devices.[21–24]

Although multi-component PVD glasses are used extensively in OLED applications, there have been few examinations of their nanoscale structure, in part because of expectations that, as a glass, the material should behave as an isotropic molecular alloy, and in part because the measurement of nanoscale structure is challenging. Multi-component organic materials can pose a challenge for many traditional thin-film characterization methods. Since the electron density of two primarily carbon-containing organic molecules varies only slightly, commonly used "hard" X-ray scattering measurements, with X-ray energies ≈10 keV, exhibit little intrinsic contrast between domains. Although techniques such as grazing incidence wide- angle X-ray scattering (GIWAXS) can be used to quantify the orientation distributions of molecular stacking planes of both components,[25] small-angle X-ray scattering (SAXS) often fails to resolve electron density modulations caused by domain-level compositional contrast. One well-established real-space technique is Scanning Transmission X-ray Microscopy (STXM), which can differentiate between components using contrast in absorption of X-rays between constituent domains.[26,27] An emerging real-space characterization that does not require a synchrotron is photo-induced force microscopy (PiFM), which uses infrared (IR) radiation in conjunction with a traditional AFM to simultaneously probe surface topography and composition.[28] Although both of these techniques are immensely powerful for materials characterization, they can suffer from the main drawbacks of all real-space techniques – probing only a limited spot size of the sample, as well as limits in xy resolution, often ≈5 nm at the best for PiFM, or 30-50 nm for carbon-edge STXM.

Resonant soft X-ray scattering (RSoXS) is a reciprocal-space measurement that scatters from organic domains with chemical bond-level specificity. RSoXS uses X-rays near the absorption edges of common organic molecules (e.g., ≈290 eV for the Carbon K-edge). When collecting patterns over a range of photon energies near an absorption edge such as the Carbon K-edge, the dominant scattering mechanism can change dramatically energy-to-energy, leading to a wealth of information not only in reciprocal space, but also hyperspectral (multiple energy) space.[29] RSoXS is a popular characterization technique for a diverse range of systems, including bulk heterojunctions,[30] polymer nanoparticles and composites,[31,32] semicrystalline semiconductors,[33] block copolymers,[34,35] and biomolecules.[36] As the technique becomes more widely employed and the number of beamlines increase, we expect the use of RSoXS to further broaden. One obstacle to conventional analytical or model-free analysis of RSoXS scattering patterns is the increased dominance of vacuum scattering due to larger ratios of roughness to thickness caused by the thin film sample form factor; advances have been made, however, in frameworks for characterizing rough films.[37,38] Here, we will expand upon the analysis framework for very rough films, widening the scope of systems that can be studied with RSoXS.

In this work, we show that co-physical vapor deposition of a mixture of the small organic molecules TPD and DO37 leads to a glass with two phases having distinct compositions, and in which the domain size and center-to-center distance of the minority phase can be tuned by choice of substrate temperature during deposition. Remarkably, even for a substrate temperature well below the glass transition temperature of both components, compositionally distinct domains with spacings of ≈50 nm are found. Both phases appear to lack long-range order and feature low free-energy domain boundaries, suggesting that the structure may be a consequence of arrested liquid-liquid phase separation, perhaps by free interface-mediated equilibration mechanisms. Using chemically sensitive photo-induced force microscopy (PiFM), we find that this phase separation closely follows the topography of the vapor-deposited film. Using resonant soft X-ray scattering (RSoXS), we identify a secondary length scale invisible to the real-space microscopy methods that we quantitatively attribute to sub-surface phase separation, indicative of a buried structure that cannot be measured by surface-sensitive methods. Using NRSS, an RSoXS pattern simulation suite, we fuse real-space imaging with composition assumptions to recreate the key scattering features and reveal the physical arrangement of secondary TPD-DO37 phase separation. In a key assertion, we show that careful treatment of surface vacuum scattering is critical for a quantitative interpretation of RSoXS. We use this structural characterization to extend the surface equilibration mechanism for PVD glass formation to a multi-component system, showing that a combination of experimental and simulated RSoXS is a powerful tool to measure otherwise invisible phase separation in multi-phase organic materials.

*Results & Discussion.*

## Co-deposited TPD:DO37 glass laterally phase separates.

Two-component glasses were vapor-deposited with a 50:50 ± 5 ratio by mass of N,N'-Bis(3-methylphenyl)-N,N'-diphenylbenzidine (TPD, glass transition temperature $T_g$ = 330 K) to Disperse Orange 37 (DO37, $T_g$ = 296 K) by mass, with structures shown in Figure 1A. The variation in ratio is calculated from slight variation in deposition rate at different lateral distances from each crucible. The two-component system will hereafter be referred to as TPD:DO37. Bulk mixtures of TPD:DO37 show two glass transitions over a wide range of compositions, including 50:50 by mass (see SI Section 1), indicating that this system forms two liquids at equilibrium. Glasses were thermally evaporated in a high vacuum chamber as illustrated schematically in Figure 1B and deposited on silicon nitride membranes, which are X-ray transparent in the energy range used for RSoXS. Samples ≈70 nm thick were vapor deposited at substrate temperatures ($T_{sub}$) from 177 K to 325 K at a rate of 0.2 nm s$^{-1}$. These conditions range from $T_{sub}/T_g$ of (0.59$T_g$ to 1.10$T_g$) for DO37, and (0.54$T_g$ to 0.98$T_g$) for TPD. All subsequent characterization was performed at ambient temperature.

AFM results, shown in Figure 1C, show that TPD:DO37 glasses deposited at higher $T_{sub}$ will phase separate with greater length scales than those deposited at lower $T_{sub}$. In the height images, shown in the top row, domains with center-to-center spacing on the order of tens of nanometers develop when the glass is deposited at 280 K and higher. All phase-separated films share the common feature of somewhat regularly-spaced "low" areas (which we will refer to as domains or holes) interspersed within a higher "matrix" area. The matrix area, in turn, has greater length-scale height fluctuations independent of these holes. Many of the AFM phase images exhibit significant contrast, suggesting lateral compositional differences. At $T_{sub}$ = 177 K, where the film is mostly flat, we see no variation in AFM phase, suggesting an evenly mixed glass. At 240 K, we see apparent compositional segregation in the AFM phase image, despite an uncorrelated AFM height image. The AFM phase images for the glasses deposited from 280 K to 325 K show systematic variations that directly correlate with the small-scale features in the topography map. Because AFM phase is sensitive to local adhesion, surface energy, and elastic modulus, these variations suggest that compositionally different phases are present in the "holes" (low regions) and the "matrix" (high regions). Additionally, the highest contrast in the AFM phase images occurs at $T_{sub}$ = 310 K, whereas at 325 K, it becomes less well-defined, despite the larger length scales. Notably, the AFM phase image for the glass deposited at $T_{sub}$ = 310 K strongly resembles those for the hexagonally perforated lamellar phase of diblock copolymers, a metastable state between hexagonal and lamellar packings.[39–41] X-ray scattering (GIWAXS) shown in SI section 3 verifies that there is no crystalline order in these films. This unusual film structure is therefore interpreted as a result of arrested liquid-liquid phase separation, perhaps including free interface-mediated equilibration mechanisms.

AFM height images of TPD:DO37 show a non-monotonic roughening then smoothing of the glasses with increasing $T_{sub}$. Figure 1D shows two-dimensional linecuts of the height images shown in the previous panel. The curves are offset for clarity. The glass deposited at 177 K is virtually flat; at the higher $T_{sub}$, both large- and small-scale height fluctuations appear, then gradually diminish at the highest $T_{sub}$. In Figure 1E, we show the RMS roughness of the linecuts over various length scales. The RMS roughness is defined as:

$$(1) \quad R_q = \sqrt{\frac{\sum (h_i - h_{ave})^2}{i}}$$

Where i is the number of pixels along the length of the linecut, $h_i$ is the thickness at pixel i, and $h_{ave}$ is the mean height of all pixels. When calculated over various length scales (i.e., varying the extents of local regions for which pixels are considered in the Equation 1 summation), we can distinguish between short-scale and long-scale roughness. Two separate trends are observed between short (50 nm to 100 nm) and long (500 nm to 2000 nm) length scales.

At small length scales, the maximum roughness is similar for the glasses deposited at $T_{sub}$ = (280 and 310) K, while at large scales, the maximum roughness definitively occurs at $T_{sub}$ = 280 K. This result suggests that there are two mechanisms of phase separation occurring. The first occurs on the micron scale, which we hypothesize to be due to an interplay between the equilibrium driving force due to surface energy and the molecular mobility, which determines to what extent the equilibrium structure can be realized before becoming trapped in the bulk. Somewhere at or between 280 K to 310 K, this effect becomes less pronounced. At the higher $T_{sub}$, we also see the effect of small-scale topographic and phase separation, where the material continues to separate into smaller, more well-organized domains. This continues to occur up to or above $T_{sub}$ = 310 K and subsides at 325 K. For both length scales, the film begins to smooth out at the highest $T_{sub}$.

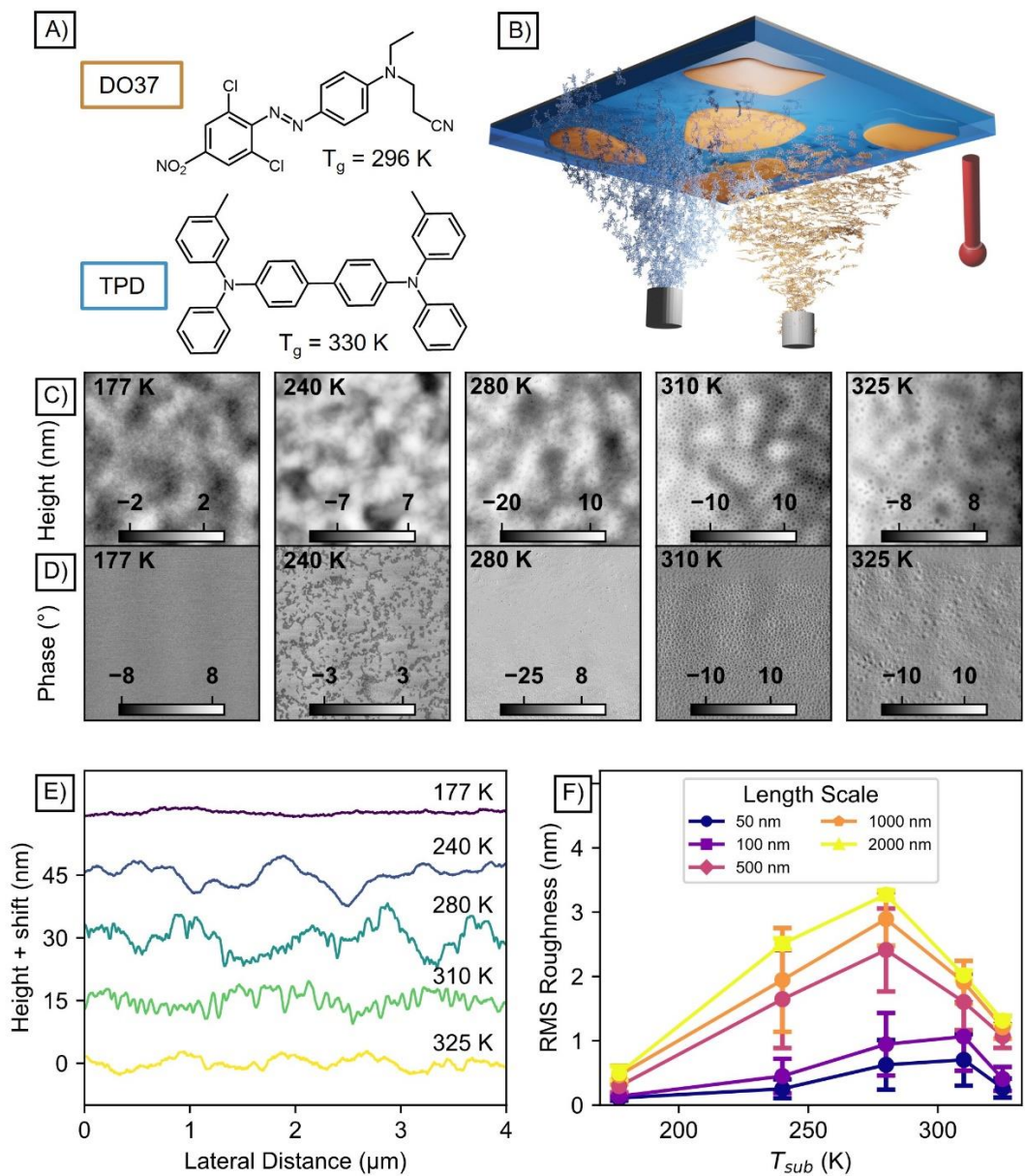

**Figure 1.** Co-deposited TPD and DO37 phase separation to an extent controlled by substrate temperature, as shown by AFM. A) Molecular structure and glass transition temperatures of TPD and DO37. B) Illustration of co-deposition scheme. C) AFM micrographs of TPD:DO37 at five $T_{sub}$. Top row shows height images with scale bars; bottom row shows corresponding phase images. All images are 4 μm x 4 μm. D) Height profiles at linear cuts along each AFM image. All are plotted on the same scale, vertically shifted for clarity, with shifts noted in the legend. E) RMS roughness values for height profiles at various length scales. Error bars are standard deviation of RMS roughness, of which there are 80 values for 50 nm, 40 for 100 nm, 8 for 500 nm, etc. Full size, higher resolution AFMs and a representative linecut are shown in SI section 2.

## PiFM shows link between phase separation and topography.

To probe the extent of the link between phase separation and topography, we use photo-induced force microscopy (PiFM) to measure the samples deposited at the highest $T_{sub}$s of 310 K and 325 K. Briefly, PiFM combines the topographical measurement of AFM with a localized infrared (IR) measurement.[42] In Figure 2, we show the topography map of the TPD:DO37 sample at $T_{sub}$ = 310 K that is taken in the same spatial region as the spectroscopic mapping; this map qualitatively matches the height profile taken by the conventional AFM shown previously. Figures 2B and 2C show the intensity of absorption at the wavenumbers indicated, (1339 and 1489) $cm^{-1}$, respectively. In Figure 2B, the regions with the greatest voltage correspond to the domains richest in DO37, while in 2C, the highest voltage occurs in TPD-rich regions. The spatial mapping confirms that the holes are rich in DO37, while the matrix areas are rich in TPD. The specific wavenumbers for mapping are picked from the IR spectra of neat TPD and DO37 shown in Figure 2D. The remaining mapping for $T_{sub}$ = 280 and 325 K, and larger versions of panels A-C, are shown in SI Section 4.

Localized IR spectra integrated from the hole-specific and matrix-specific regions of interest reveal that both areas contain both chemical species, in different ratios depending upon $T_{sub}$. Figure 2D shows spectra of the holes and matrix of the samples deposited at 310 and 325 K, along with neat spectra of TPD and DO37 for reference. We highlight the distinguishing peaks for DO37 at (1136, 1150, and 1339) cm$^{-1}$ and TPD at 1489 cm$^{-1}$ with vertical dashed lines. The TPD peak has been reported to correspond to the C-C vibration,[43] while the peak at 1339 cm$^{-1}$ for DO37 may correspond to a C-N stretch as in an aromatic amine.[44] It is first clear that neither the holes nor the matrix of either sample fully match the neat spectra for TPD or DO37; instead, they appear to be a combination of the two. If the domains were homogenously mixed, the spectra for the two regions should be identical to those for the neat films; to quantify the difference between spectra, we subtract them, as shown in Figure 2E. Particularly focusing on the peaks at 1339 cm$^{-1}$ and 1489 cm$^{-1}$, we see the greatest difference in spectra for the film deposited at 325 K. We therefore conclude that the film deposited at 325 K phase separates to a greater extent than that at 310 K.

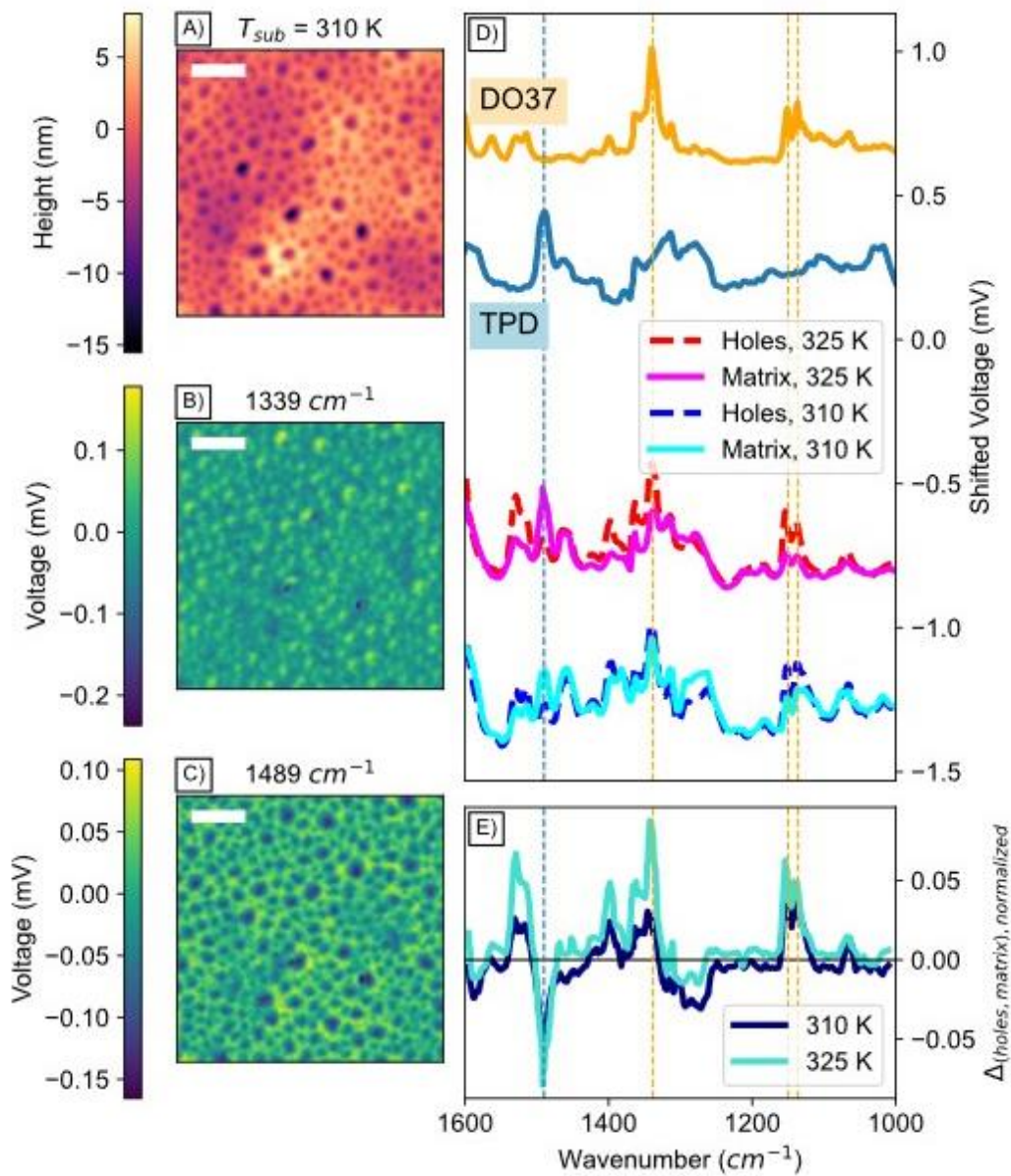

*Figure 2*. PiFM shows variable phase separation along with topography for TPD:DO37. A) Height image of sample deposited at 310 K. B) $T_{sub}$ = 310 K sample measured at 1339 cm$^{-1}$ and C) 1489 cm$^{-1}$, corresponding to DO37 and TPD content at maximum voltage, respectively. D) PiFM IR spectra of neat components and multi-component films. The top curve, orange, shows the spectrum for a neat film of DO37. Prominent peaks at 1339, 1150, and 1136 cm$^{-1}$ are shown as orange, dotted vertical lines. The curve below, in solid blue, is that for neat TPD, with the prominent peak at 1489 cm$^{-1}$ shown as a blue vertical dotted line through all spectra. The bottom two pairs of spectra show the spectra in the low, or "hole" region as a dashed line, paired with a solid line for the high, or "matrix" region, for the samples deposited at the two highest substrate temperatures.

E) Normalized difference between hole and matrix IR spectra for glasses deposited at the two substrate temperatures shown. Scale bars for microscopy images are 300 nm.

While the PiFM provides qualitative information, its limited spatial resolution and depth sensitivity prevent a fully quantitative understanding of the phase purity. In SI Section 5, we show the hole and matrix spectra of both samples alongside a variable weighting of the neat spectra. From a rough analysis focusing on the peak structure around 1339 cm$^{-1}$, we estimate the "hole" domains of the sample at 310 K to be 40%/60% TPD to DO37, while a 30%/70% ratio for $T_{sub}$ = 325 K. While the ratio is more distinct in the higher $T_{sub}$ sample, we use caution assigning a definite number to our results. One limitation of PiFM is its spatial resolution; although it is a rapidly improving technique, its maximum spatial resolution is ≈5 nm. While this is likely sufficient for the higher $T_{sub}$ glasses with average domain sizes of (90 and 130) nm, this begins to become uncertain for the glass deposited at 280 K, which is why we do not attempt a quantitative interpretation here. A second limitation of PiFM is its uncertain depth sensitivity. The technique likely probes the top ≈10 nm of the film, and therefore investigates only the surface of the ≈70 nm-thick films without sensitivity to bulk or buried structure. If the laterally phase-separated film structure is assumed to be uniform through the entire thickness, then the PiFM image could be interpreted as describing the whole film structure, but PiFM alone is not sufficient to validate that assumption.

**RSoXS reveals phase separation at multiple length scales.**

Soft X-ray scattering confirms the phase separation and morphologies seen with AFM. In Figure 3 we show the scattering intensity versus the scattering vector, q, for each PVD glass measured at two photon energies. Panel A shows data measured at 270.5 eV, well below the carbon K-edge, to enhance contrast sensitivity to surface vacuum scattering. Panel B shows data measured at 285.0 eV, which is the photon energy of maximum contrast between TPD and DO37. At both energies, the higher-q peaks (and shoulder, for $T_{sub}$ = 240 K) for the four $T_{sub}$ are commensurate with the trend expected from AFM imaging. The peak/shoulder positions, from low to high $T_{sub}$, at q ≈ (0.15, 0.01, 0.07, and 0.05) $nm^{-1}$ correspond to average center-to-center spacings of ≈ (40, 60, 90, and 130) nm, consistent with AFM images. The peak position as a function of $T_{sub}$ is presented as filled circles in Figure 3C. The shoulders at low q ≈ 0.005 $nm^{-1}$, correspond to a length scale of ≈ 1 μm that is consistent with the large-scale film height fluctuations; this peak position remains relatively invariant among all $T_{sub}$.

At 285.0 eV, the scattering intensity is significantly enhanced, and a third feature emerges in the I vs. q pattern as shown in Figure 3B. Most notably in the sample deposited at 280 K, a peak appears at q ≈ 0.03 $nm^{-1}$ (d ≈ 210 nm). At the other substrate temperatures, a subtle shoulder with approximately the same temperature dependence as the higher q peak appears, with positions summarized in Figure 3C. The medium-q peak positions are estimated from the point at which the first derivative of I vs. q corresponds to zero. This scattering feature corresponds to an intermediate size scale that is larger than the domain size, but smaller than the micron-scale height fluctuations. There are no obvious features in the AFM that correspond to these intermediate length scales, so we speculate that it corresponds to a secondary, buried phase separation. The origin of this peak is further explored later in the manuscript, using a data fusion scattering simulation that supports this speculation. For the measurements taken at 285.0 eV, a fluorescence background appears at high q, which we will subtract in subsequent analysis.

We interpret the energy dependence of the TPD:DO37 RSoXS patterns through a calculation of the scattering contrasts shown in Figure 3D. Near-edge X-ray absorption fluorescence spectroscopy (NEXAFS) measurements of neat films were converted to optical constants δ and β (SI Section 6), where the X-ray refractive index is n = 1 – δ + iβ, through a Kramers-Kronig transformation. Contrast is an intrinsically binary concept, and scattering contrast between two materials can be calculated by equation 2:[38]

$$(2) \qquad C = E^4[(\delta_A(E) - \delta_B(E))^2 + (\beta_A(E) - \beta_B(E))^2]$$

Where A and B refer to the different materials (including vacuum, where δ and β are equal to 0). This, in the TPD:DO37:vacuum system, there are $_3C_2$ = 3!/((3-2)!2!) = 3 binary contrasts as shown in Figure 3D. At energies above and below the carbon absorption edge, hydrocarbon-vacuum contrast is enhanced; therefore, we see a greater contribution of scattering from features caused by whole-film height fluctuations. If there is TPD-DO37 phase separation, its scattering signal will be obscured by that from the vacuum contrast, especially if it is on the same length scale as the topographical features. However, by measuring at a resonant energy, we observe the additional medium-q features due to the enhanced TPD-DO37 contrast. These peaks are not visible or are greatly diminished at 270.5 eV, suggesting that they are due to a structural feature other than the roughness. There is no energy at which TPD-DO37 contrast overtakes the hydrocarbon-vacuum contrasts; therefore, no energy can be assigned to one specific feature; multiple mechanisms must be considered at all energies. Of note, the medium-q resonant peak appears the most separated at $T_{sub}$ = 280 K, which showed the largest amount of RMS roughness in the AFM data at long length scales; as it decreases at the higher $T_{sub}$, the peaks become less pronounced.

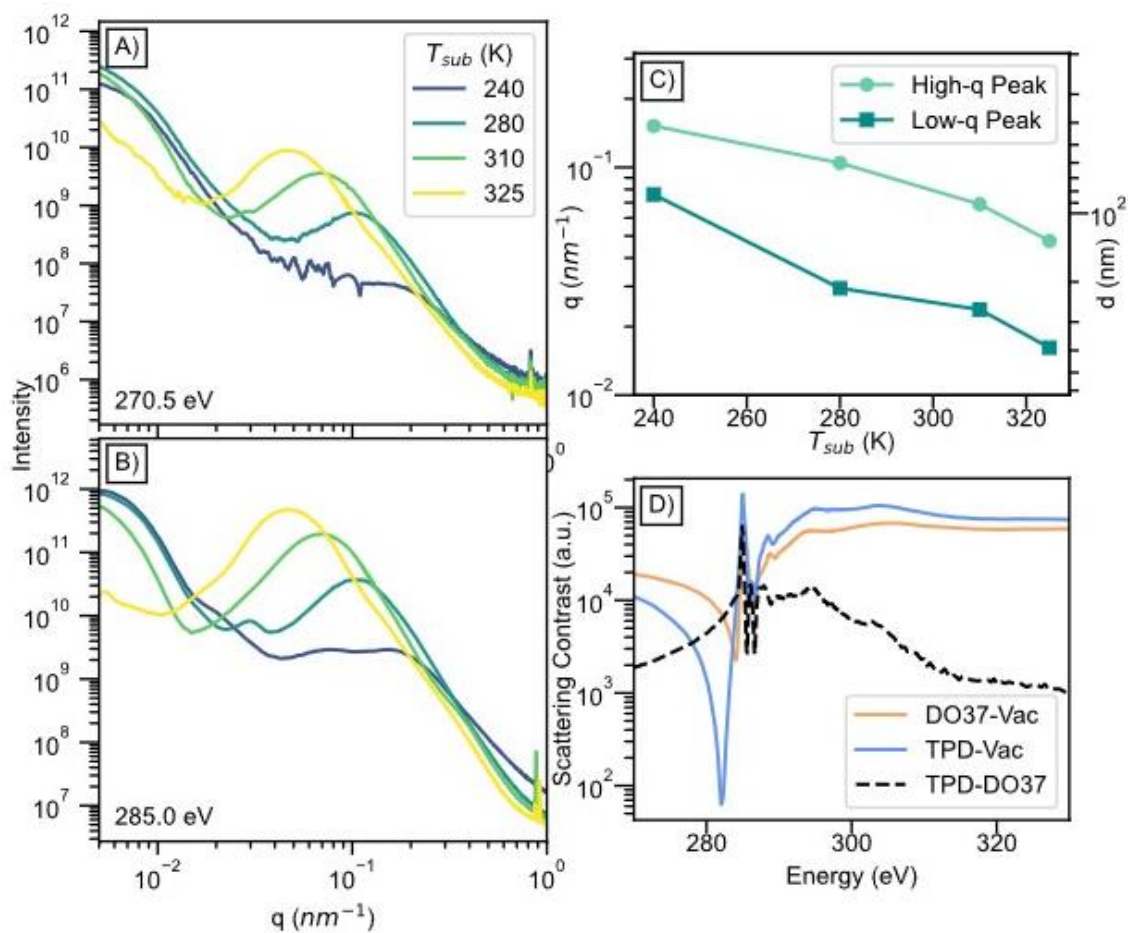

**Figure 3.** Resonant Soft X-ray scattering shows phase separation at both non-resonant and resonant energies for glasses deposited at $T_{sub}$ = 240 K and above. A) Scattering at the pre-edge photon energy of 270.5 eV. Noise around q ≈ 0.1 nm$^{-1}$ is due to stitching at low signal at the high end of the SAXS and low end of WAXS data. B) Scattering patterns at 285.0 eV. A second peak or faint shoulder appears at lower q values than the primary peaks at q ≈ (0.04 to 0.1) nm$^{-1}$. C) Peak positions of low-q and high-q peaks taken at E = 285.0 eV as a function of substrate temperature during deposition. Peak position decreases with increasing $T_{sub}$, indicating that length scales become large at higher $T_{sub}$. Error bars are calculated by minimum and maximum error possible in sample-to-detector distance. They are smaller than the symbol size. D) Hydrocarbon-vacuum and TPD-DO37 contrasts as a function of energy, calculated from magic-angle NEXAFS of both compounds.

**Emergent length scale revealed with enhanced scattering contrast.**

As shown in Figure 3, a medium-q peak emerges at the resonant energy of 285.0 eV that is absent or greatly diminished in the scattering pattern at 270.5 eV. In Figure 4, we show the normalized, fluorescence-subtracted I vs. q pattern at energies below and near the carbon absorption edge for the glasses deposited at all four $T_{sub}$ in panels A-D. For all films, the low-q and high-q peaks exist at all energies, both resonant and non-resonant (except for a few cases for $T_{sub}$ = 240 K, which will be discussed subsequently). The peaks grow in intensity as the energy increases, reaching a maximum at E = 285.0 eV, and decrease with further increases in energy. For the resonant energies, from approximately (284.5 to 286.0) eV, the medium-q peaks appear. Finally, we note that in the high-q WAXS range (q > 0.4 nm$^{-1}$), a small shoulder appears in the scattering, most noticeably for the sample deposited at 310 K. This could be due to shorter range orientational correlations; however, this is beyond the scope of the current study.

For the glass deposited at $T_{sub}$ = 240 K, scattering at the resonant energies has substantially more features compared to the scattering at non-resonant energies, shown in 4A. For example, the scattering pattern at 285.25 eV has peaks at ≈ (0.025, 0.07, and 0.2) nm$^{-1}$ that are absent from the scattering at 286.75 eV. Notably, the $T_{sub}$ = 240 K film is the only one that has a curve absent of a peak at 286.75 eV, suggesting that any small-scale features are due completely to TPD-DO37 contrast. As can be seen by the combination of height and phase image in Figure 1, this glass is the only one in which the phase separation and topography are not correlated. Due to a lack of meaningful real-space imaging, we do not investigate this sample further, but the experimental data make a case for a hierarchical structure with several length scales.

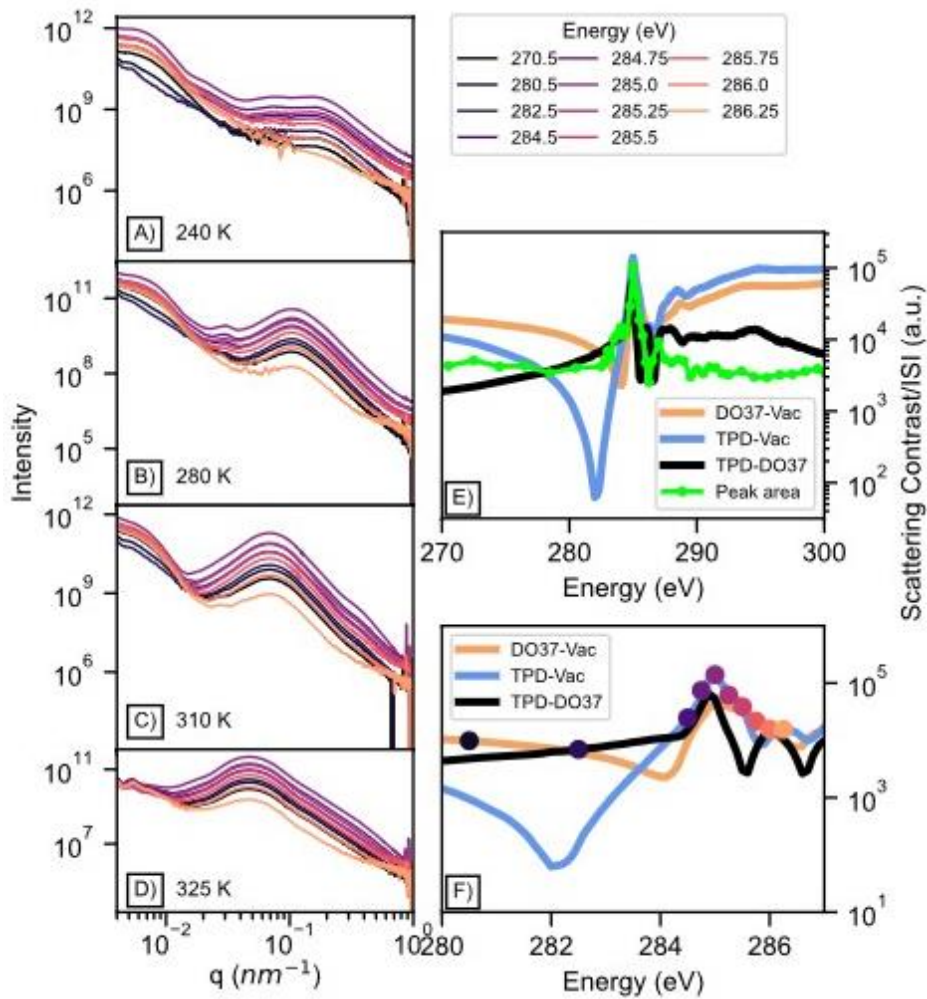

**Figure 4**. Energy-dependent X-ray scattering patterns show additional peaks at resonant energies. A) through D) show intensity vs. scattering vector q for glasses deposited at $T_{sub}$ = (240, 280, 310, and 325) K. Scattering is shown for the energy range in the pre-edge of the carbon K- absorption edge. E) Total integrated scattering intensity of background-subtracted peak at q = (0.02 to 0.07) nm$^{-1}$ for the glass deposited at $T_{sub}$ = 280 K, as a function of energy. Background subtraction procedure is shown in SI Section 7. For reference, the calculated TPD-DO37 and hydrocarbon-vacuum contrasts shown in Figure 3D are plotted for comparison. F) The calculated scattering contrast at each energy (270.5 eV not shown) for each scattering pattern shown in panels A-D. The color of each symbol matches the corresponding scattering pattern in the preceding panels. The calculated scattering contrasts from Figures 3D and 4E are once again shown for reference; each symbol is plotted on the line which corresponds to the highest calculated contrast at that energy.

To examine the origin of the emergent medium-q peak, we plot the energy dependence of its intensity in Figure 4E. When compared to the calculated scattering contrasts (shown in Figure 3D), the energy dependence of the emergent peak most resembles the TPD-DO37 scattering contrast. At energies below the peak at 285.25 eV, there is no strong local minimum as would be expected in both calculated hydrocarbon-vacuum contrasts (and indeed most hydrocarbon-vacuum contrasts, as such a local minimum occurs where $\beta$ is small at photon energies lower than resonant and where $\delta$ crosses zero as it changes sign, as it does in most hydrocarbons). At higher energies above the step edge, there is less agreement in the fit, but the intensity of the emergent medium-q peak still agrees more strongly with TPD-DO37 contrast than hydrocarbon-vacuum contrast because it does not increase with energy and reach a plateau above the step edge. This behavior is a common feature of hydrocarbon-vacuum contrast in most systems, since $\delta$ is relatively flat and $\beta$ increases up to the step edge intensity, where it plateaus. The lack of perfect agreement could be attributed to even slight energy calibration issues: in the event that either spectrum were off by even 0.1 eV, that could significantly change the exact shape of the contrast function. Alternatively, artifacts in the NEXAFS spectrum such as subtle differences in post-edge slope decay caused by electron yield physics could compromise the calculated binary contrasts. These considerations underscore the importance of careful, well-calibrated NEXAFS to RSoXS data analysis, and highlight challenges in measuring the complex index of refraction indirectly through NEXAFS electron yield modalities. The peak intensity here is derived from a fit of the peak in which we subtract all background contributions to the scattering and integrate only the area of the peak, as shown in SI Section 7. We choose to do this for the $T_{sub} = 280$ K film only, since the peak is completely separable from all others. The reasonably good agreement between the intensity of this medium-q feature and the expected TPD-DO37 binary contrast strongly indicates that the medium-q feature arises from compositional heterogeneity via the phase separation mechanism described previously. We did not observe phase separation on these length scales in the AFM or PiFM.

The energies at which RSoXS measurements were collected are highlighted with respect to the calculated scattering contrasts in Figure 4F. Over the range of energies shown in this figure, at most energies the scattering contrasts vary by only a factor of 1.5 to 3 relative to one another, meaning all scattering patterns have significant contributions from at least two binary contrasts. This result shows that it is important to carefully consider multiple binary contrast contributions when drawing conclusions about the origin of scattering patterns in RSoXS; there are almost never specific energy choices that will completely isolate composition (hydrocarbon-hydrocarbon) heterogeneity from roughness (hydrocarbon-vacuum).

**Q-bounded Integrated Scattering Intensity quantifies scattering at distinct length scales.**

Total scattering intensity or total scattering invariant (TSI) is a concept that is often used to quantify phase purity in multicomponent organic systems.[45] The standard TSI is calculated by equation 3,

$$(3) \quad TSI = \int_{q=0}^{q=\infty} I(q) \cdot q^2 dq$$

which is a measure of the total scattering invariant; notably for TSI to be correct requires integration across all reciprocal space frequencies. Equation 3 also assumes that the structure is isotropic, having frequencies in all three spatial dimensions that are similar to the dimension(s) measured. Here, we use a q-bounded integrated scattering intensity ("ISI") to partially isolate the origins of scattering intensity arising from features of different length scales. Functionally, this modifies equation 3 by changing the limits of the integral. Results are shown in Figure 5 for three ranges in q, labeled "low", "medium", and "high". Exact integration limits are given in the Table

1. For low q, all samples are integrated at the same region of the scattering vector q, as the large length scale is functionally invariant for all $T_{sub}$. The limits for intermediate q were chosen to coincide with the maximum in intensity for the $T_{sub} = 280$ K sample, and the maximum in the second derivative for the shoulder in the samples deposited at other $T_{sub}$. Finally, the limits for high q are set such that the center of the range is at the center of the peak for the highest-q feature. We stress that these are mostly qualitative decisions, and the following analysis is not rigorously quantitative.

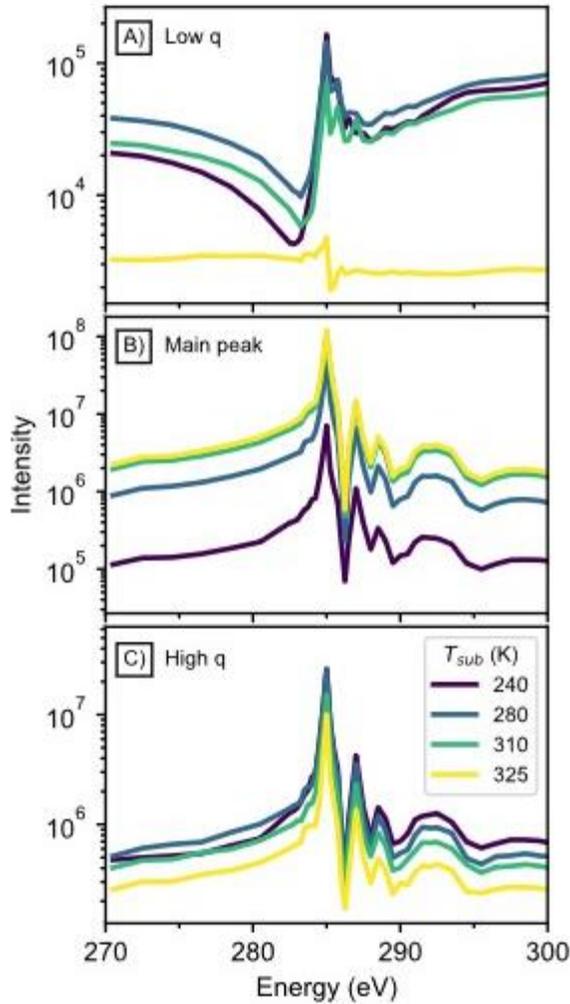

**Figure 5**. Integrated scattering intensities $\int_{q_1}^{q_2} I(q) * q^2 dq$ for the three different peak regions at each $T_{sub}$. Different q-regions are used for each $T_{sub}$ to capture region of interest. The limits for the q-ranges are summarized in Table 1. A) ISIs at the low-q region. B) ISIs at the intermediate-q region. As the size of the q-range is not uniform at intermediate q, intensities are scaled accordingly for uniformity. C) ISIs at the high-q region.

| $T_{sub}$ (K) | Low (nm$^{-1}$) | Medium (nm$^{-1}$) | High (nm$^{-1}$) | All (nm$^{-1}$) | Emergent (nm$^{-1}$) | Medium Range (nm$^{-1}$) |
|---|---|---|---|---|---|---|
| 240 | 0.003, 0.010 | 0.050, 0.200 | 0.200, 1.000 | 0.003, 1.000 | N/A | 0.15 |
| 280 | 0.003, 0.010 | 0.050, 0.200 | 0.200, 1.000 | 0.003, 1.000 | 0.023, 0.040 | 0.15 |
| 310 | 0.003, 0.010 | 0.040, 0.150 | 0.200, 1.000 | 0.003, 1.000 | 0.015, 0.030 | 0.11 |
| 325 | 0.003, 0.010 | 0.025, 0.100 | 0.200, 1.000 | 0.003, 1.000 | N/A | 0.075 |

**Table 1.** Integration limits for curves shown in Figure 5.

The low-q scattering shows a difference in dominant mechanism between the $T_{sub}$ = 325 K sample and the three others. Figure 5A shows that the samples deposited at $T_{sub}$ = (240 to 310) K all have an ISI vs. photon energy dependence that looks largely similar to the hydrocarbon-vacuum contrasts shown in Figure 3D. Most notably, the dependence has a local minimum at energies around (282 to 283) eV at the δ zero-crossing mentioned earlier and a relatively high plateau above the step edge to the β post-edge plateau. Finally, the dominant contrast mechanism for the film deposited at 325 K appears not only faint, but also consistent with the TPD-DO37 contrast due to the positive correlation between intensity and energy at lower energy. In view of the comparably mild height fluctuations in the highest $T_{sub}$ sample (seen in Figure 1), it is reasonable that there would be little contribution of material-vacuum contrast to this signal. The faint material contrast-like dependence suggests that there is a small degree of large-scale phase separation in this film. An alternative hypothesis is that, in the $T_{sub}$ = 325 K sample, the length scale of these height fluctuations results in a feature that is outside of the q-range of the detector; therefore, we do not capture the full behavior of the scattering caused by the feature.

The energy dependences of both the medium-q (5B) and high-q (5C) ISI are substantially different from that of the low-q ISI. Both the medium and high q energy dependences appear roughly consistent with the calculated TPD-DO37 contrast, missing the aforementioned pre-edge δ zero-crossing and post-edge β plateau influences. The principal difference in ISI vs. photon energy among the samples appears to be an overall trend in magnitude with substrate temperature. For the medium-q scattering, the overall intensity increases with increasing substrate temperature, while at low-q, intensity decreases with substrate temperature. This suggests that glasses deposited at lower substrate temperatures have more TPD-DO37 scattering at smaller length scales, while those deposited at higher substrate temperatures (presumably where there is more mobility) have TPD-DO37 separation on larger length scales. However, we highlight the results of our analysis to urge caution in this simplistic interpretation. The ISI in all q-ranges includes some contributions from both TPD-DO37 and hydrocarbon-vacuum contrasts. Therefore, phase purity should only be measured after careful treatment of vacuum scattering contributions. We cannot specify a lower bound on what constitutes a "rough" film for general RSoXS analysis because contrasts vary enormously system-to-system depending on spectroscopic details. Furthermore, the *scattering factors* of both hydrocarbon-hydrocarbon and hydrocarbon-vacuum structures can influence their relative contributions within finite q-limits. The results shown in Figure 5 illustrate that we can qualitatively determine purity trends for a sample, and perhaps determine a q-range in which it is appropriate to make a purity determination. To further separate the sources of scattering at different length scales and contrast mechanisms, we turn to RSoXS pattern simulation.

# Computational reconstructions of thin-film nanostructure illustrate origins of scattering features.

The analysis in the preceding section demonstrates that it is important to consider multiple contrast mechanisms that contribute to an RSoXS dataset, especially in rough films. Here, we will show a computational modelling framework that can be used to understand the origin of the multiple scattering features. By using the NIST RSoXS Simulation Suite (NRSS) and its component software CyRSoXS,[46] we are able to simulate scattering from virtual morphologies built using high-quality real-space AFM imaging as a basis for model construction.

We use the AFM height images to construct separate simplified models that simulate scattering from different contrast mechanisms. We design three models to isolate 1) large length-scale hydrocarbon-vacuum scattering, and 2) small length-scale hydrocarbon-vacuum scattering, as well as 3) TPD-DO37 scattering from any contrasting regions. We provide a full description of the model in SI section 8 and will explain the key features here. For each representation, we construct a 1x2048x2048 voxel film, in which each voxel is a cube with a side length of 1.95 nm (determined by the AFM resolution). The real-space map for the voxels was found from image processing using the Python package scikit-image. Each voxel contains a specified volume fraction ($V_{frac}$) of TPD, DO37, and vacuum. For the simulated X-rays to interact realistically with the virtual material representation, the TPD and DO37 are assigned material-specific energy-dependent isotropic optical constants ($\delta$, $\beta$) as described above and shown in SI Section 6.

The AFM height images (example Figure 6A) are used to construct the first two simplified models in which scattering only results from roughness (surface vacuum scattering), with scattering from TPD-DO37 contrast purposefully removed by keeping phase composition uniform throughout. The two simplified models are denoted "nm-scale roughness" (Figure 6B and C) and "µm-scale roughness" (6D and E). The volume fractions shown are for vacuum, while $V_{frac, TPD}$ = $V_{frac, DO37}$ = $(1 - V_{frac,vac})/2$, resulting in an equally mixed material of TPD and DO37. For the µm-scale representation, a scikit-image Gaussian filter with a large standard deviation was applied to the height image to eliminate small-scale height variations. To create the model for the nm-scale, the same Gaussian-filtered image was subtracted from the original image, leaving only small-scale height variations. For the nm-scale height model, additional vacuum was added uniformly such that the total volume fraction of vacuum was equal to that in the µm-scale representation. Since the TPD and DO37 are evenly mixed, this image processing leaves a virtual scattering model in which the only optical constant variations are caused by inclusion of vacuum in each voxel. Scattering intensity is proportional to $\Delta n_{AB}^2$ for two components in a scattering medium; in this case, anywhere where there is vacuum, the refractive index is modified, increasing the scattering contrast between high and low regions.

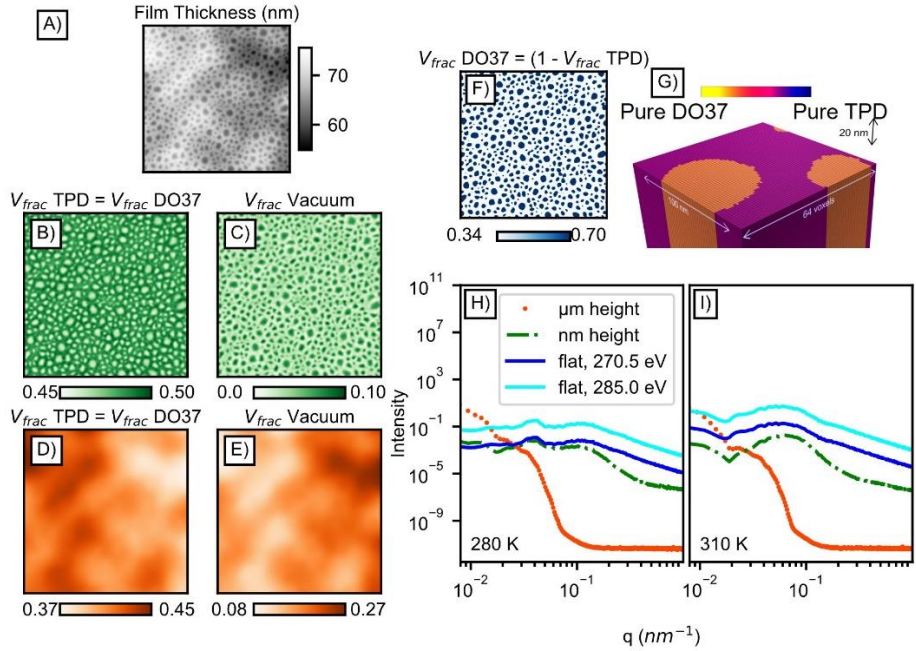

**Figure 6**. AFM Height and Phase images are used to "deconstruct" real samples into three simplified models which isolate and reproduce the key scattering mechanisms. All images shown are 500x500nm. A) AFM height and image for film deposited at 310 K. B) Volume fractions of TPD and DO37 (equal) and C) vacuum for "nm-scale roughness" model. D) Volume fractions of TPD and DO37 and E) vacuum for "µm-scale roughness" model. F) Volume fractions of TPD for the "flat material" model, along with a 3D representation of the entire model shown in G). H) and I) I vs. q scattering patterns from simulated scattering from the three models shown at two X-ray energies for $T_{sub}$ = (280 and 310) K, respectively.

A third simplified model shown in Figure 6F is generated using the AFM height image to assign composition, rather than height. The layer that we construct is effectively flat, as illustrated in Figure 6G, with $V_{frac,vac} = 0$ everywhere. We use the height image to binarize the 2D image into "hole" domains and "matrix" areas. Because the hole domains occupy only 4% to 8% of the surface area of the 2D image, this assignment cannot be reconciled with pure TPD and DO37 phases that were both initially deposited with roughly 0.5 whole-film volume fraction. We achieved mass balance by pinning the hole domains at a fixed fraction of DO37 (70% in the models shown in Figure 6), which was estimated from the PiFM results. The volume fractions of TPD and DO37 in the "matrix" were assigned to maintain 50/50 volume balance between DO37 and TPD in the film. This assumption significantly informs our understanding of the surface equilibration mechanism, as discussed later.

By simulating scattering from the three simplified models, the origin of different scattering features can be separated. Figures 6H and 6I show the isolation of the three different scattering mechanisms for both $T_{sub}$ = (280 and 310) K, at both the non-resonant energy of 270.0 eV and the resonant energy of 284.5 eV. For the "nm-scale height" and "µm-scale height" traces, the pattern shapes are the same at both energies and are therefore only plotted once. The scattering patterns that result from the flat material are shown for both (270.0 and 284.5) eV. At these two energies, the hydrocarbon-vacuum contrast is roughly the same. However, the TPD-DO37 contrast varies greatly between the two energies, which results in the ≈2 order of magnitude difference in intensity between the two blue curves. The variation in these curves with energy, combined with the invariance of the nm- and µm-scale height curves, would result in a curve with changing shape if the layers were to be simply added together to construct a rudimentary 3-layer model, showing that the appropriate ratios of relative scattering are critical to reconstructing an experimental scattering pattern.

**Vacuum-scattering is a dominant mechanism in rough films and must be accounted for when quantifying phase purity with RSoXS.**

The results shown in Figure 6 demonstrate a significant contribution of hydrocarbon-vacuum scattering at most energies, including resonant energies. Caution is therefore strongly advised for the commonly employed RSoXS relative phase purity measurement, which is a model-free analysis framework especially common in OPV blends that compares relative TSI at resonant energies as a representation of material phase purity. This approach likely convolutes hydrocarbon-vacuum contrast with hydrocarbon-hydrocarbon contrast, with unrecognized influences from sample roughness affecting the phase purity conclusions. Here, we have shown a framework which may be used to correctly quantify the hydrocarbon-vacuum contrast contributions from AFM height images and the measured optical constants of all materials in the film. From this, photon energies and q-ranges at which vacuum makes significant contributions can be quantitatively avoided during a TSI-purity analysis. This strengthens the conventional standard of "how smooth" a film needs to be for RSoXS; the final answer depends upon the length scales of the roughness, the optical constants of the materials involved, and the energy of the measurement, making the answer unique for each film.

In the present work, we make frequent use of a q-bounded ISI, which is justified based upon the results in Figure 6. At all values of q, both TPD-DO37 and hydrocarbon-vacuum scattering make a significant contribution to the total scattering. For example, in the region from $q \approx (0.007$ to $0.01)$ $nm^{-1}$, the hydrocarbon-vacuum scattering is approximately 3 orders of magnitude greater than TPD-DO37; in other words, no information about domain purity could be found from this measurement. At higher q values, such around the main peaks at $q \approx 0.1$ $nm^{-1}$, the TPD-DO37 and hydrocarbon-vacuum scattering may be roughly equal at 270.0 eV, while the TPD-DO37 scattering dominates at resonant energies. The results shown in Figure 6 are not a full model in and of themselves; rather, they are separate simplified models to illustrate different contributions, so we now consider a full descriptive, quantitative model informed by our results.

**Constructing a multiple length scale-informed scattering model.**

We now consider a comprehensive model that includes both realistic roughness and compositional phase separation. The key features of the model that are used to reproduce the scattering data are the AFM roughness, a domain identification from the AFM, and an assumption of 50/50 volume balance and how it is distributed throughout the film. Three-dimensional illustrations of the final models are shown in Figure 7A and B for two substrate temperatures. We use the AFM height images shown in Figure 1 to binarize the 2D image into "hole" domains and "matrix" areas. In the same procedure as described for the split model illustration in Figure 6, we keep material balance in the film by adding DO37 to the high areas while keeping the composition within the holes constant.

Figure 7C and 7D show the agreement between the scattering pattern of our model for the glasses deposited at (280 and 310) K, respectively. These two $T_{sub}$s were chosen due to the quality of the AFM phase images and the medium-q "emergent peak" behavior. Experimental data (reproduced from Figure 4) is shown in the inset, with the same color-coding. Behavior at $q < 0.006$ nm$^{-1}$ is not shown for the simulation results due to artifacts caused by finite effects. The simulation reproduces key features in the experimental data, most notably the peak positions in q and the energy dependence of the scattering, especially at low q. Both the simulated and experimental data show peaks at $q \approx$ (0.035 and 0.1) nm$^{-1}$ for the sample deposited at 280 K, and at $q \approx 0.008$ nm$^{-1}$ for the $T_{sub} = 310$ K film, along with a hint of a shoulder at $q \sim 0.025$ nm$^{-1}$. We point out that the simulated data does appear to have features at $q > 0.2$ nm$^{-1}$ that is inconsistent with the experimental data, suggesting that we may be inadvertently introducing features due to finite simulation or voxel size.

The energy dependence of ISI vs. energy across two separate q-ranges is reproduced from the model. Figure 7E and 7F show data for I vs E that is integrated at low q (7E) and in the range of the main peak (7F) for both $T_{sub}$. The exact integration bounds are the same as used for the experimental data in Figure 5, and are given in Table 1. The experimental data (reproduced from Figure 5) is shown in the insets. Qualitatively, we see that the energy dependence of the scattering between low q and the main peak q is extremely different, as seen in experiment. Most significantly, we see the expected local minimum at energies below 284 eV, with a sharp increase at 284 eV followed by a slight decrease up to 290 eV, finishing with the expected post-edge plateau at higher energies with broad features at about (294 and 305) eV. For the ISI vs. photon energy of the main q peak, we see qualitative agreement that is different from the low q energy dependence, and moderately consistent with the medium q. We hypothesize that a potential source for this disagreement is any small error in the measured dielectric constants. Within the energy range of disagreement, the NEXAFS absorption spectra (shown in the SI) for both TPD and DO37 show many features, frequently crossing over one another. Therefore, even a slight energy calibration issue in one spectrum could lead to disagreement in this highly featured region.

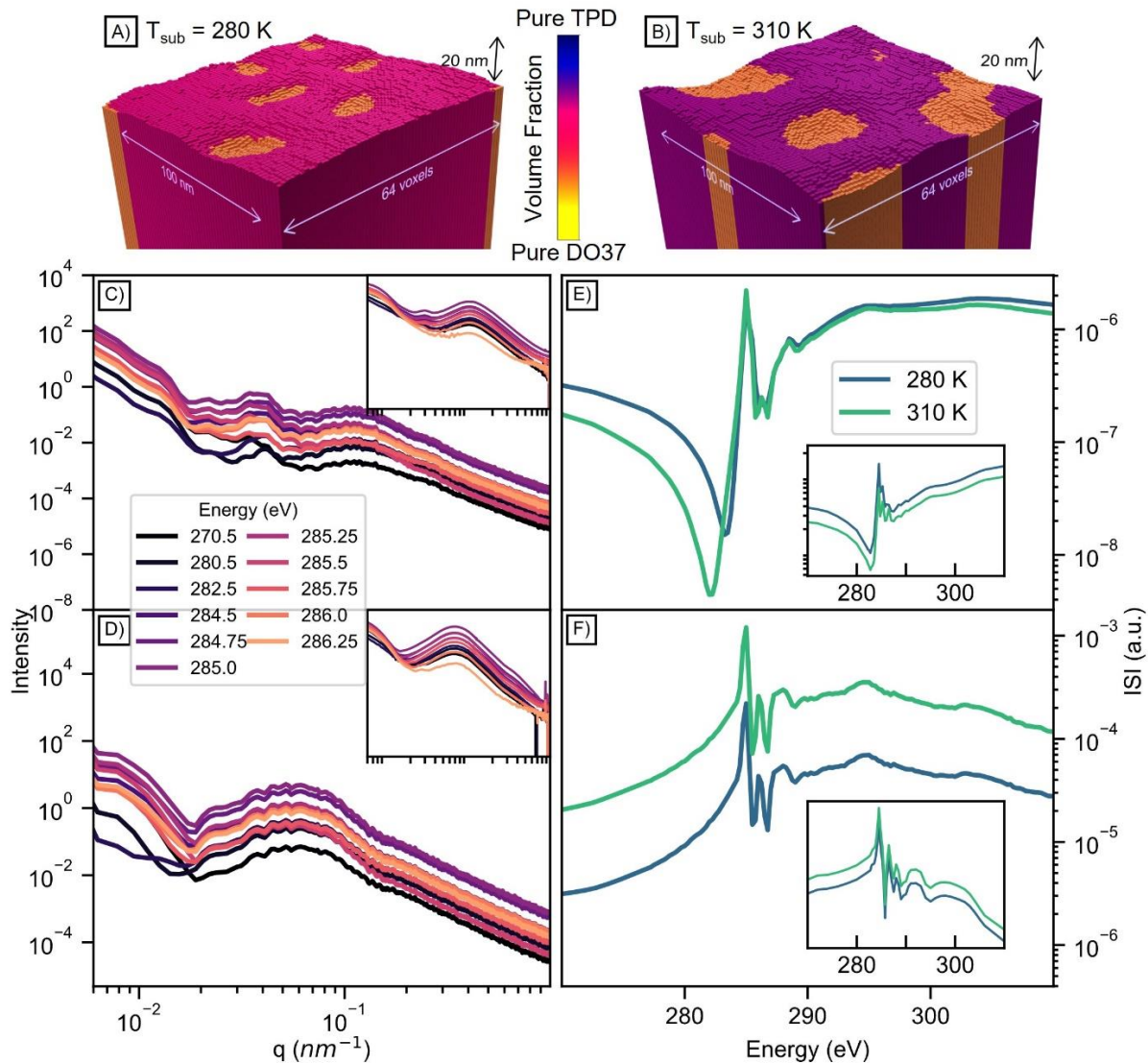

**Figure 7.** A model of film morphology used to simulate scattering in NRSS yields I vs q and I vs E profiles that are qualitatively consistent with experimental data. A) Illustration of volume fractions of TPD, DO37, and vacuum in the models for 280 K and B) 310 K. C) I vs. q scattering patterns for simulated data from the 280 K model at various energies. Experimental data is shown in the inset for the same q-range; for labels, see Figure 4. D) Scattering patterns for the 310 K model, with experimental data shown in inset. Energy color labeling is the same as shown in panel C. E) and F) Integrated $I(q)q^2$ simulation data for morphologies modeled from films deposited at (280 and 310) K, respectively. Data shown in E) is integrated from q = 0.007 to 0.01 nm$^{-1}$, consistent with the experimental data. Lower q data is not included due to artefacts from the simulation method. Scattering intensity in F) is integrated at ((0.050 to 0.200) and (0.040 to 0.150)) nm$^{-1}$ to be consistent with experimental integration. Corresponding experimental data is shown in inset. Data in all insets is shown in greater detail in Figures 4 and 5, where initially presented.

## Higher substrate temperatures lead to phase separation with longer center-to-center length scales.

For the five films deposited in the present work at constant deposition rate, the center-to-center distances between domains increased at increasing substrate temperatures. We assume that phase separation is thermodynamically favorable at all substrate temperatures. As surface mobility increases with temperature, the two components are able to separate at increasingly large length scales. For the glass deposited at 177 K, $0.54T_{g,TPD}$, there is no composition variation. We attribute this result to low molecular mobility at this substrate temperature; when deposited as a single component at these temperatures, TPD shows little to no thermal stability or enhanced density,[47] and would be expected to have a very low surface diffusion coefficient of $10^{-44}$ $m^2$ $s^{-1}$ based on extrapolated data of Zhang et. al.[48] We thus conclude that the TPD:DO37 deposited at 177K represents a non-equilibrium single-phase blend. On the other hand, glasses deposited at high $T_{sub}$, such as 310 K, appear to phase separate as a consequence of immiscibility coupled with high surface mobility. Despite being evenly mixed when first vapor-deposited (similar, we would expect, to the $T_{sub}$ = 177 K glass), DO37 should be highly mobile above its glass transition temperature, and TPD has been shown to form stable glasses at this $T_{sub}$, indicating that it has sufficient mobility to rearrange at the surface. From a qualitative perspective, the most revealing sample is the $T_{sub}$ = 240 K glass. Despite both components being at least 50 K below their (neat) glass transition temperatures, the center-to-center distance for phase separation is greater than 20 nm, which could not be achieved without significant lateral mobility of the deposited molecules after initial contact with the substrate. This result confirms an essential feature of the surface equilibration mechanism: that rapid equilibration occurs at the surface during physical vapor deposition, and then this structure is trapped by subsequent deposition.

Generally, we understand the increasing length scales with higher substrate temperature to be a consequence of higher mobility at the increased $T_{sub}$. This is roughly consistent with the "Rate Temperature Superposition" principle (see, for example, Bishop et. al.[20]), which equates the anisotropic molecular orientation that results from raising the substrate temperature during deposition to depositing more slowly in single-component vapor-deposited glasses. Here, we have not tested the dependence of the separation on the deposition rate, so we cannot comment on the rigorous validity of RTS for this system; this could be a fruitful area for further study.

Although there are many examples of co-deposited systems showing phase separation,[49] to our knowledge this is the first direct characterization of phase separation in a completely amorphous system. As a result, we cannot draw general conclusions about phase separation for glassy structures, but we expect the following features to be important. The intermolecular interactions between the two components determine the thermodynamic driving force for phase separation; greater incompatibility is expected to yield larger length scales of separation. In addition, intermolecular interactions have been shown to have a significant effect upon surface diffusion coefficients[50] in single-component systems, and may well have an impact for co-deposited systems. Finally, DO37 and TPD have fairly different glass transition temperatures – (296 and 330) K, respectively. Given the crucial role of surface mobility, systems with different ratios of $T_g$ will exhibit different features in terms of phase separation.

**Relationship between phase separation and topography.**

On at least one length scale, the TPD-DO37 separation follows the topography of the vapor-deposited film, as evidenced by the significant peak around q ~ (0.05 to 0.2) nm$^{-1}$ (referred to as the "medium-q" peak) which appears at both resonant and non-resonant energies. At resonant energies, it is significantly enhanced, indicating close correlation between composition and height fluctuations. This assertion is further confirmed by the PiFM data, which finds distinct compositions in the low and high areas. The data in Figure 5 shows that the intensity of the scattering at this length scale depends on the energy in a fashion consistent with that matching the TPD-DO37 contrast, suggesting that the phase separation dominates the signal at this length scale. In the sample deposited at the lowest $T_{sub}$ of 177 K, there is neither material nor height fluctuations throughout the film; however, upon moving to the next highest temperature of 240 K, fluctuations in both appear, suggesting that the two evolve in a concerted fashion. Despite this correlation, we find that there is a secondary material length scale that does not correlate with height fluctuations.

The appearance of the medium-q emergent peak, especially in the 240 K and 280 K samples, shows that there is a secondary phase separation length scale that is not apparent from the AFM or PiFM alone. As the emergent peak appears at lower q, it is evidence of a longer length scale. It does not appear at a second harmonic of the main peak, indicating it is a distinct structural feature. Interestingly, as can be seen in Figure 5A, the intensity vs. energy dependence of the scattering at low q for the film deposited at 325 K is much more similar to TPD-DO37 contrast than hydrocarbon-vacuum, unlike the other three substrate temperatures. This suggests that at this highest substrate temperature, which is almost 20 K above the glass transition temperature of pure DO37 and only 5 K below that of pure TPD, there is enough mobility to achieve more than 100 nm of separation between phases.

**NRSS Model construction provides a large space to explore structural features that are signatures of the surface equilibration mechanism.**

The physically-informed purity assumptions used to model the film in NRSS result in a reasonably good agreement that reproduces the essential features in the experimental data – namely, the peak positions and energy dependence. The optimal results in the simulated film, with regards to purity and length scales of phase separation, can be used to inform our understanding of the surface equilibration mechanism in the co-deposited TPD and DO37. To reproduce experimental scattering while keeping mass balance, we pin the "low" domains to a composition that is 70% DO37, roughly estimated from the PiFM information. We balanced the mass to 50/50, as known from deposition, by adding DO37 uniformly to the high areas. An alternative procedure, in which we grew the majority-DO37 domains to achieve 50/50 mass balance, and then filled the majority TPD and majority DO37 domains with a reciprocal amount of material, led to simulated scattering that did not resemble experiment. One possible explanation is that the local environment of a surface molecule influences its surface diffusion coefficient. As the higher-$T_g$ component, a molecule of TPD should have lower mobility at any deposition temperature. In a 50/50 mix of TPD and DO37, it will have enhanced mobility; upon reaching a region locally rich in TPD (perhaps templated by the film below), the surface diffusion coefficient will decrease, growing the size of the region.

We stress that the example shown in Figure 7 is just a starting point for a full quantitative fit of domain purity, composition gradients, and even molecular orientation in these films. It is difficult to believe that the current system has perfectly sharp interfaces. It also likely has at least some degree of orientation in- or out-of-plane, since vapor-deposited glasses are rarely isotropic,[51] or at interfaces between domains.[33] A fit to all of these parameters simultaneously is quite possible, but it is outside of the scope of the current work. Strategies to approach this could be reverse Monte Carlo[52] or genetic algorithm methods,[53] and should be pursued in the future to improve these models. Therefore, we present the current system as a starting point and hope that the model can provide a starting point for more advanced structural studies.

## Conclusions.

In this work, we have shown how a two-component vapor-deposited glass phase separates to a degree that is determined by the substrate temperature during deposition. High quality real-space imaging combined with RSoXS reveal both material and height fluctuations in the vapor-deposited films. Using a model-free analysis based on RSoXS contrast fundamentals, we can analyze the dependence of scattering intensity on both reciprocal-space scattering vector and X-ray energy to deduce the operative scattering mechanism (due to height or material fluctuations) at each length scale. We further use the real-space imaging to build a model using the NRSS X-ray scattering simulator. We demonstrate "splitting" the real-space images into simplified models of their constituent structural features to separate the origin of different scattering features. Finally, we fuse the data together with reasonably informed assumptions and generate comprehensive real-space models that capture the essential features of the RSoXS experiment. We demonstrate that this method can be used to understand the surface equilibration mechanism in multi-component vapor-deposited glasses. An improved understanding through these studies could be used to engineer materials with finely engineered degrees of phase separation, which could be useful in OPVs, photonics, or patterned device fabrication.

## Materials and Methods

TPD (Sigma-Aldrich) and DO37 (BOC Sciences) were used as-received. Glasses were vapor-deposited in a high vacuum chamber with a base pressure of $10^{-5}$ Pa as detailed in previous publications[54] and illustrated schematically in Figure 1B. The TPD and DO37 were heated in separate crucibles through a resistive wire. Glasses were deposited on Silicon Nitride Membranes (Norcada) affixed to liquid nitrogen-cooled copper fingers controlled by proportional-integral-derivative (PID) heaters (Lakeshore). Samples were vapor deposited at substrate temperatures ($T_{sub}$) from 177 K to 325 K at a rate of 0.2 nm s$^{-1}$ with the rate monitored by quartz crystal microbalance (QCM, Sycon). Before depositing, the two molecules were brought to the desired rate while the sample was covered by a shield; the shield was removed once the desired rate was achieved.

The glasses deposited at $T_{sub}$ = 240 K to 325 K were measured at beamline SST-1 of the NSLS-II at Brookhaven National Laboratory.[55] All were measured in transmission geometry with both SAXS and WAXS detectors, using approximate sample-to-detector distances of 500 and 38 mm, respectively. All data shown in this publication is taken at the carbon edge, ranging from 270 to 340 eV, with an energy resolution of 0.2 eV. Data was reduced using the PyHyperScattering Python package.[56] SAXS and WAXS data from the same run is stitched together, as shown in Figure 3, with small gaps in data caused by cutoffs in accessible data range.

NEXAFS data was taken at SST-1 at the NEXAFS endstation. Data for TPD was taken at the RSoXS end-station in TEY on a 50-nm neat film vapor deposited at 270 K. DO37 NEXAFS was taken on crystalline powder at the NEXAFS endstation. Spectra were corrected and fit in QANT[57] using previously published procedures. QANT-corrected NEXAFS was then used in kkcalc[58] to generate X-ray optical constants.

AFM was collected on a Bruker Dimension Icon in tapping mode. Scans used for modeling were taken at a resolution of 2048 lines with 2048 pixels per line, corresponding to a lateral resolution of 1.95 nm. Images were processed with pySPM.[59] PiFM images were collected using a Molecular Vista One AFM and NCH-PtIR PiFM cantilevers (Molecular Vista). Nanoscale infrared images were acquired by tuning an infrared quantum cascade laser to a specific wavenumber and acquiring 256 x 256 pixel images at 0.1 lines/s. PiFM images were flattened to remove line-to-line variations in intensity.

X-ray scattering was simulated using NRSS, an open-source RSoXS scattering simulation package.[46,60] Models for scattering were 2-dimensional arrays of 1.95 nm voxels, comprising TPD, DO37, and vacuum volume fractions determined by the AFM. Data was reduced in PyHyperscattering.

## Acknowledgements.

This research was funded by the National Institute of Standards and Technology. C.B. and T.J.F. acknowledge funding from the National Research Council Research Associateship Program.

M.E.F. and M.D.E. were supported by the U.S. Department of Energy, Office of Basic Energy Science, Division of Materials Science and Engineering Award DE-SC0002161. C.G.B. acknowledges support from the National Science Foundation under Award CHE-2304613. This research used beamline SST-1 of the National Synchrotron Lightsource II, a U.S. Department of Energy (DOE) Office of Science User Facility operated for the DOE Office of Science by Brookhaven National Laboratory under Contract No. DE-SC0012704.

## Author Contributions.

D.M.D., M.D.E., T.J.F., and M.E.F. conceived of the project and designed the experiments. M.E.F. produced the samples. C.E.B., T.J.F., and C.G.B. performed measurements. C.E.B. performed the data analysis and simulations and prepared the manuscript. E.G. and C.J. aided with experiments at NSLS-II at Brookhaven National Laboratory. All authors read and approved the final manuscript.

## Notes.

The authors declare no competing financial interest.